\documentclass[reprint, amsmath,amssymb, aps, prb, superscriptaddress]{revtex4-2}
\usepackage{graphicx}
\usepackage{dcolumn}
\usepackage{bm}
\usepackage[colorlinks=true, linkcolor=blue, urlcolor=blue, citecolor=blue]{hyperref}
\usepackage{dcolumn}
\usepackage{cleveref}
\usepackage{color}
\usepackage{multirow}
\usepackage{float}
\usepackage{url}
\usepackage{physics}
\usepackage{subfigure}
\usepackage[utf8]{inputenc}
\usepackage{siunitx}
\usepackage{ragged2e}
\usepackage{makecell}
\usepackage{fontawesome}
\usepackage{titlesec}
\usepackage[T1]{fontenc}
\usepackage[dvipsnames]{xcolor}
\usepackage[version=3]{mhchem}
\usepackage[normalem]{ulem}
\graphicspath{ {./Figures/} }

\begin{document}

\title{Fractionation as a Tool to Control Mn/Al Site Mixing and Magnetism in \ce{CaMn_{2+$x$}Al_{10-$x$}}}

\author{S. Kr\'{o}lak}
\email{szymon.krolak@pg.edu.pl}
\affiliation{Ames National Laboratory, U.S. Department of Energy, Iowa State University, Ames, Iowa 50011, USA}
\affiliation{Department of Physics and Astronomy, Iowa State University, Ames, Iowa 50011, USA}
\affiliation{Faculty of Applied Physics and Mathematics and Advanced Materials Center, Gdansk University of Technology, Narutowicza 11/12, 80-233 Gdansk, Poland}

\author{S. Kumari}
\affiliation{Ames National Laboratory, U.S. Department of Energy, Iowa State University, Ames, Iowa 50011, USA}
\affiliation{Department of Physics and Astronomy, Iowa State University, Ames, Iowa 50011, USA}

\author{S. L. Bud’ko }
\affiliation{Ames National Laboratory, U.S. Department of Energy, Iowa State University, Ames, Iowa 50011, USA}
\affiliation{Department of Physics and Astronomy, Iowa State University, Ames, Iowa 50011, USA}

\author{P. C. Canfield}
\email{canfield@ameslab.gov}
\affiliation{Ames National Laboratory, U.S. Department of Energy, Iowa State University, Ames, Iowa 50011, USA}
\affiliation{Department of Physics and Astronomy, Iowa State University, Ames, Iowa 50011, USA}


\begin{abstract}
We report single crystal growth and physical property characterization of \ce{CaMn_{2+$x$}Al_{10-$x$}} crystals grown using a fractionation approach. Three consecutive batches were obtained from the same initial melt composition, systematically sampling neighboring, narrow regions of the ternary Ca-Mn-Al phase diagram. Single-crystal X-ray diffraction measurements reveal Mn/Al site mixing, with the excess Mn concentration, $x$, decreasing systematically with decreasing decanting temperature. A low-temperature upturn in magnetization and a maximum in electrical resistivity both become progressively weaker with decreasing excess Mn content, suggesting that the previously reported magnetic properties attributed to stoichiometric \ce{CaMn2Al10} instead arise from an inhomogeneous distribution of excess Mn atoms. More broadly, the results presented in this work demonstrate how fractionation can help distinguish intrinsic magnetic behavior from disorder-induced effects associated with an inhomogeneous distribution of magnetic ions in materials with a finite width of formation, particularly when itinerant magnetism is under consideration.

\end{abstract}

\maketitle


\section{INTRODUCTION}

Fragile magnets \cite{Canfield_fragile} are magnetic systems in which the ordering temperature can be brought sufficiently low that fluctuations associated with the suppression of magnetic order significantly affect the physical properties \cite{Taufour2016, Paglione2003, Xiang2021, Das2026, Krellner2011, Custers2003, Gegenwart2008, Brando2016}, making them model systems for studying quantum-critical behavior. One approach to identifying candidate fragile magnetic systems is to search for materials showing intrinsically small saturation/ordered moments. However, the magnetic properties of such systems can be difficult to characterize unambiguously. Unlike rare-earth-based systems, in which local moments provide a relatively well-defined magnetic entropy scale \cite{Canfield_fragile, Budko1999, Krlak2022, Garcia}, itinerant magnets often exhibit only weak thermodynamic signatures \cite{ZrZn2, ZrZn2_HC, Sc3In, Sc3In_HC, TiAu, Ti3Cu4, La5Co2Ge3, La4Co4X}. For this reason, establishing whether observed magnetic properties are intrinsic can be particularly challenging in itinerant magnets and has, in some cases, led to extrinsic magnetic signals being incorrectly attributed as an intrinsic property of the studied material \cite{Coey2005}.

From a crystal-growth perspective, fractionation provides a useful approach to addressing the above problem. Introduced by Slade and Canfield \cite{Slade2022}, it is particularly valuable for multinary crystal growth when the relevant region of the phase diagram is poorly known or unknown. Detailed descriptions of the method are given in Refs. \cite{Slade2022, Canfield2016}; here, we briefly summarize the basic concept. In fractionation, the remaining liquid from one growth (decant) is reused as charge in a subsequent growth, allowing several consecutive cuts through the phase diagram to be obtained from a single starting composition. This approach can be particularly useful for studies of itinerant magnetic systems, as separating the crystal-growth process into successive temperature intervals can distinguish the growth of the target phase from that of (potentially magnetic) impurity phases \cite{Huyan2024, Schmidt2025}, which could otherwise dominate the magnetic response. Moreover, when the target phase exhibits a finite homogeneity range, fractionation allows crystal growth to be restricted to a selected temperature--and thus composition--interval, which can help reduce the influence of compositional variation on the measured properties.

\ce{CaMn_{2+$x$}Al_{10-$x$}} was previously reported to be stoichiometric, i.e. \ce{CaMn2Al10}, with a small effective moment $\mu_{\mathrm{eff}} = 0.83,\mu_{\mathrm{B}}$/Mn and magnetic entropy associated with the low-temperature specific-heat feature equal to only $\approx9\%$ of $R\ln 2$ \cite{Steinke2015}. It exhibited maxima in low-temperature AC magnetic susceptibility, electrical resistivity and specific heat. Together with the observation that the real part of the AC susceptibility follows $\chi' \sim T^{-1.2}$, these features were interpreted as an indication of incipient ferromagnetism, with a Curie temperature $T_C < 2$ K. In this work, we use the fractionation approach to grow single crystals of \ce{CaMn_{2+$x$}Al_{10-$x$}}. The growth yielded three consecutive batches of \ce{CaMn_{2+$x$}Al_{10-$x$}} single crystals with systematically decreasing excess Mn ($x$) concentration, as determined by single-crystal X-ray diffraction. The magnetic and transport properties change systematically with composition: the low-temperature upturn in magnetic susceptibility and the maximum in electrical resistivity, both of which were considered signatures of the incipient ferromagnetism, become progressively weaker as $x$ decreases. These trends allow us to reassess the previously proposed intrinsic magnetic behavior of stoichiometric \ce{CaMn2Al10} and, more broadly, demonstrate the usefulness of fractionation for separating weak, intrinsic magnetic responses from those associated with magnetic disorder resulting from non-stoichiometry.


\section{EXPERIMENTAL METHODS}
\subsection{Single crystal growth}

Single crystals of \ce{CaMn_{2+$x$}Al_{10-$x$}} were grown from Al flux using Al shot (Alfa Aesar, 99.999\%), Mn chunks (ACI Alloys, 99.99\%, metal basis), and Ca chunks (Minteq, 99.9+\%) in a Ca:Mn:Al molar ratio of 6:11:83. Al and Mn were loaded into a 2 ml Canfield Crucible Set (CCS) \cite{Canfield2016, CCS_page}, after which Ca was added inside an Ar-filled glovebox. To minimize Ca oxidation during necking/sealing, the open end of the fused-silica tube was closed with Parafilm\textsuperscript{\tiny{\faRegistered}}. Before necking, small holes were punctured in the film to allow heated Ar gas to escape. For flame sealing, the tube was backfilled with $\approx$ 1/3 atm Ar. 

\begin{figure}[!ht]
\centering 
\includegraphics[width=\columnwidth]{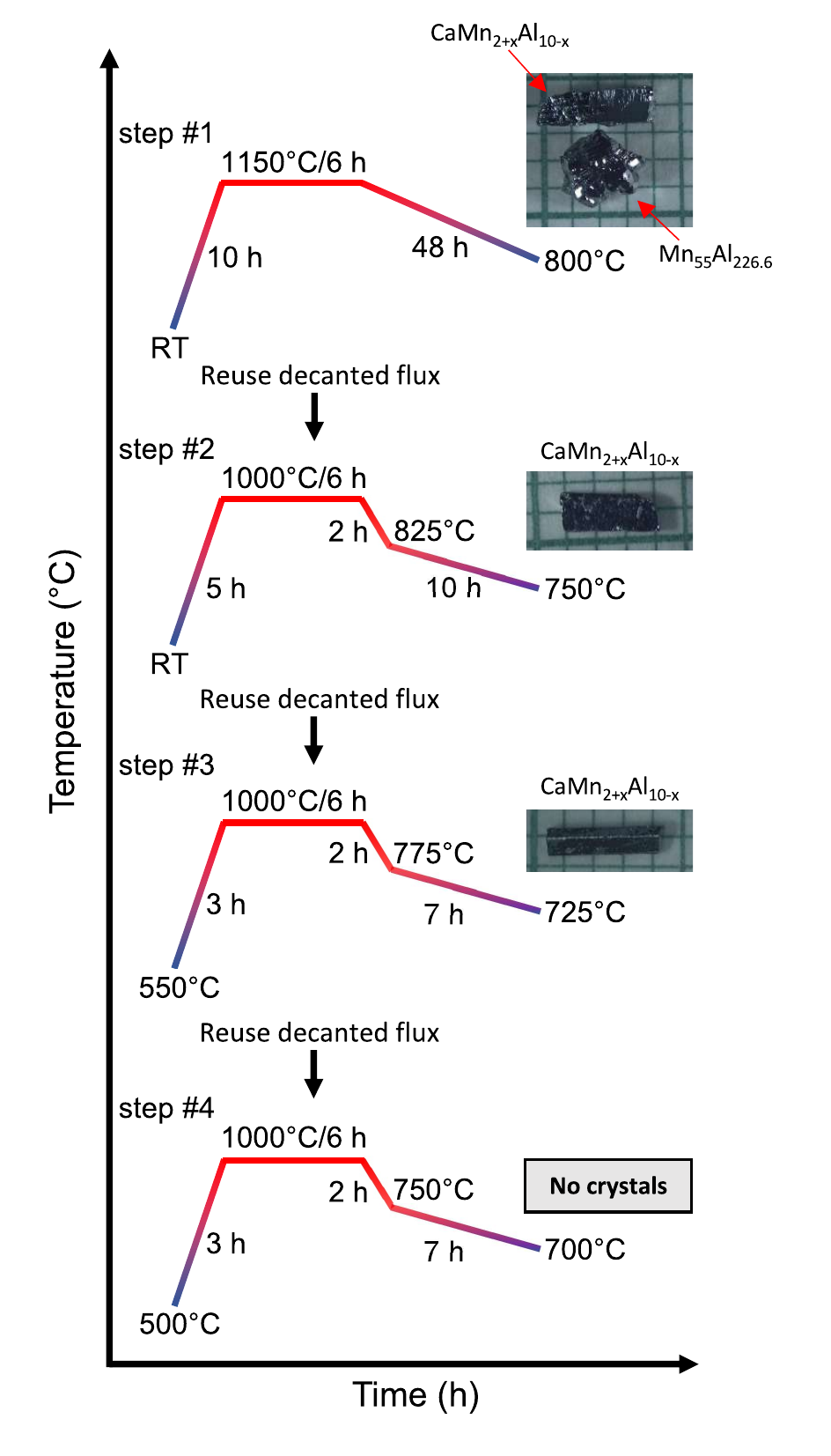}
\caption{ 
Schematic of the fractionation procedure used to grow \ce{CaMn_{2+$x$}Al_{10-$x$}} single crystals.
Following each growth step, the remaining flux was decanted and used as a starting material for the subsequent growth. Representative photographs of the obtained crystals are shown for steps \#1-3. The initial growth over a broad temperature range yielded two phases, \ce{Mn_{55}Al_{226.6}} and \ce{CaMn_{2+$x$}Al_{10-$x$}}, while steps \#2 and \#3 produced only \ce{CaMn_{2+$x$}Al_{10-$x$}} crystals. Step \#4 resulted in almost all of the material decanting to the catch-side, with no sizable single crystals observed on the growth-side.
}
\label{Fig:fractionation}
\end{figure}

Unfortunately, in Ref. \cite{Steinke2015} the only growth information that was given was, "Single crystals of \ce{CaMn2Al10} were grown from self-flux...". This is clearly inadequate information for reproduction of the experiment, lacking information about initial stoichiometry as well as any temperature profile information. As such, then, we needed to treat this growth as a new experimental synthesis rather than a reproduction of prior work. Here, a fractionation approach \cite{Slade2022, Huyan2024, Huyan2023} was utilized, with subsequent growth steps presented in \Cref{Fig:fractionation}. The first logical step, i.e., heating to the highest accessible temperature and decanting, was omitted; a different batch with the same initial composition, \ce{Ca6Mn11Al83}, was heated to \SI{1150}{\degreeCelsius}, kept there for 6~h and decanted at \SI{1000}{\degreeCelsius}. No crystals were grown and all of the material decanted to the catch-side, indicating that \ce{Ca6Mn11Al83} forms a single-phase liquid at \SI{1000}{\degreeCelsius}. In step \#1, the CCS was heated to \SI{1150}{\degreeCelsius}, kept there for 6h, and cooled to \SI{800}{\degreeCelsius} over 48h, after which the tube was removed from the furnace and excess flux was decanted with a centrifuge. This step resulted in two crystal morphologies: thick rods and thick hexagonal tablets. Elemental analysis indicated that the rods were the target phase, \ce{CaMn_{2+$x$}Al_{10-$x$}} \cite{Steinke2015}, while the elemental composition of the hexagonal tablets corresponded most closely to \ce{Mn_{55}Al_{226.6}}, also known as \ce{MnAl_{4.12}} \cite{Shoemaker1989}, which was additionally confirmed with powder X-ray diffraction (pXRD), see \Cref{Fig:Mn55_pXRD} in Supporting Information. The thick rods of \ce{CaMn_{2+$x$}Al_{10-$x$}} were found to grow on (and out of) the surface of the \ce{Mn_{55}Al_{226.6}} crystals, with \ce{Mn_{55}Al_{226.6}} found mostly at the bottom of the crucible, indicating that \ce{CaMn_{2+$x$}Al_{10-$x$}} was probably a secondary solidification for this initial growth step. Thus, with approximately half of the starting load recovered as decant (0.7641~g out of 1.6919~g), we proceeded to step \#2 of the fractionation. Because the decanting temperature in the first step (\SI{800}{\degreeCelsius}) was substantially lower than the maximum temperature (\SI{1150}{\degreeCelsius}), the CCS was heated to \SI{1000}{\degreeCelsius} in step \#2, rather than to \SI{1150}{\degreeCelsius}. The temperature was then lowered to \SI{25}{\degreeCelsius} above the previous decanting temperature, and crystals were grown by slowly cooling the melt. This procedure yielded \ce{CaMn_{2+$x$}Al_{10-$x$}} as the only crystalline phase, while, again, leaving a measurable amount of decant. The procedure was therefore repeated in step \#3 and again yielded \ce{CaMn_{2+$x$}Al_{10-$x$}}. In the final, step \#4, almost all of the material decanted to the catch-side, with no sizable single crystals obtained on the growth-side.

\subsection{Physical property characterization}

To ensure consistency in the structure-property analysis, all measurements at each fractionation step were performed on material obtained from the same single crystal. Prior to measurements, single crystals were etched with dilute NaOH and HCl solutions to remove residual Al flux from the surface. Etching did not significantly affect the intrinsic physical properties of the crystals; see \Cref{Fig:etching} in the Supporting Information.

The crystal structure was characterized by single-crystal X-ray diffraction (SC-XRD) and powder X-ray diffraction (pXRD). SC-XRD measurements were performed at room temperature ($\approx 300$ K) using a Rigaku XtaLab Synergy-S diffractometer with Ag radiation ($\lambda = 0.56087$ \AA) in transmission geometry, operated at 65 kV and 0.67 mA. Single crystals were mounted on nylon loops using vacuum grease. Data-collection strategies were generated with CrysAlisPro (Rigaku OD, 2023), which was also used for data integration and reduction. Numerical absorption corrections were applied by Gaussian integration over face-indexed crystal models. Structures were solved by intrinsic phasing with SHELXT and refined with SHELXL. pXRD measurements were performed using a Rigaku MiniFlex II diffractometer with Cu-K$\alpha$ radiation ($\lambda = 1.5406$ \AA), and the obtained data were refined using the Rietveld method \cite{Rietveld}, as implemented in GSAS-II \cite{GSAS} software package.

Chemical composition was studied by energy-dispersive X-ray spectroscopy (EDS) using a JCM-7000 NeoScope\textsuperscript{\tiny{\faTrademark}} benchtop SEM. Thick rods of \ce{CaMn_{2+$x$}Al_{10-$x$}} were cut perpendicular to the rod axis using a wire saw, and the cross sections were polished.

DC magnetization was measured as a function of temperature (1.8 -- 350 K) and magnetic field (up to 5.5 T) using Quantum Design Superconducting Quantum Interference Device (SQUID) magnetometers. Thick rods of \ce{CaMn_{2+$x$}Al_{10-$x$}}, with the rod axis coinciding with the crystallographic $c$-axis, were aligned with the rod axis parallel to the applied field by mounting them between two plastic half-straws inserted into a full straw. Temperature-dependent magnetization was measured in the linear $M(H)$ regime ($\mu_\mathrm{0}$H = 0.1 T), thus the magnetic susceptibility was defined as $\chi=M/H$.

Electrical resistivity was measured using the DC resistivity option of a Quantum Design DynaCool Physical Property Measurement System (PPMS), with 8 mA current in a four-probe geometry. The current was applied along the crystallographic $c$-axis (rod axis), with the magnetic field perpendicular to the $c$-axis. Electrical contacts were made by attaching 50-$\mu$m Pt wires to etched crystal surfaces using two-component EPO-TEK H20E silver epoxy, followed by curing at \SI{120}{\degreeCelsius} for 1h. The resistance of each contact was below 2 $\Omega$.


\section{RESULTS}


\subsection{SC-XRD and EDS}

\renewcommand{\arraystretch}{1.15}
\begin{table*}[t]
\caption{Single-crystal X-ray diffraction refinement data, atomic positions,
site occupancies, and equivalent isotropic displacement parameters for
\ce{CaMn_{2+$x$}Al_{10-$x$}} crystals grown in steps \#1--3. Data were collected at room
temperature using Ag radiation ($\lambda = 0.56087$~\AA). $U_{\rm eq}$ is defined as one-third of the trace of the
orthogonalized $U_{ij}$ tensor \cite{Fischer1988}. Unphysical occupancy of the Al3 site is underlined (see text for details).
}
\label{tab:SC-XRD-occ}

\begin{ruledtabular}
\resizebox{\textwidth}{!}{%
\begin{tabular}{
    ll
    @{\hspace{1.1em}\vrule width 1.1pt\hspace{1.1em}}
    lcccccc
}


\multicolumn{9}{c}{
Step \#1
} \\

\hline

Space group; $Z$
& P4/nmm; 4
& Atom & Wyckoff & Occupancy & $x$ & $y$ & $z$
& $U_{\rm eq}$ (\AA$^2$) \\

\cline{3-9}

$a$ (\AA)
& 12.8189(3)
& Mn1 & $8i$ & 0.987(10) 
& 0 & 0.24237(2) & 0.25351(2) & 0.00627(5) \\

$c$ (\AA)
& 5.1445(2)
& Ca1 & $2a$ & 1.007(11) 
& 0 & 0 & 0 & 0.00861(10) \\

Volume (\AA$^3$)
& 845.37(5)
& Ca2 & $2c$ & 0.995(11) 
& $\frac{1}{2}$ & 0 & 0.47989(7) & 0.00794(10) \\

No. of parameters
& 49 
& Al1 & $8g$ & 0.992(11) 
& 0.17434(2) & 0.17434(2) & 0 & 0.00963(10) \\

$\theta$ range (deg)
& 1.773--31.800
& Al2 & $8j$ & 0.992(11)
& 0.89052(2) & 0.39052(2) & 0.03150(6) & 0.00946(9) \\

Refl. collected; independent refl.
& 23002; 1554
& Al3 & $8i$ & \underline{1.153(12)} 
& 0 & 0.25831(2) & 0.75103(4) & 0.00667(8) \\

Goodness of fit
& 1.054 
& Al4 & $8h$ & 0.990(11) 
& 0.11343(2) & 0.11343(2) & $\frac{1}{2}$ & 0.00885(9) \\

$R_1$; $wR_2$ [$I>2\sigma(I)$]
& 0.0299; 0.0392 
& Al5 & $8j$ & 0.995(11) 
& 0.82590(2) & 0.32590(2) & 0.48734(6) & 0.01028(10) \\

Diffraction peak/hole
(e$^{-}$/\AA$^3$)
& 0.529; $-0.465$ & \multicolumn{7}{c}{} \\


\hline
\hline

\multicolumn{9}{c}{
Step \#2
} \\

\hline

Space group; $Z$
& P4/nmm; 4
& Atom & Wyckoff & Occupancy & $x$ & $y$ & $z$
& $U_{\rm eq}$ (\AA$^2$) \\

\cline{3-9}

$a$ (\AA)
& 12.8664(4)
& Mn1 & $8i$ & 0.998(8) 
& 0 & 0.24109(2) & 0.25436(2) & 0.00559(4) \\

$c$ (\AA)
& 5.1453(2)
& Ca1 & $2a$ & 1.000(8) 
& 0 & 0 & 0 & 0.00870(8) \\

Volume (\AA$^3$)
& 851.77(6)
& Ca2 & $2c$ & 1.005(8) 
& $\frac{1}{2}$ & 0 & 0.47559(5) & 0.00745(8) \\

No. of parameters
& 49
& Al1 & $8g$ & 1.001(8) 
& 0.17529(2) & 0.17529(2) & 0 & 0.00857(7) \\

$\theta$ range (deg)
& 1.766--29.209
& Al2 & $8j$ & 1.000(8)
& 0.89166(2) & 0.39166(2) & 0.03659(4) & 0.00879(7) \\

Refl. collected; independent refl.
& 12001; 1181
& Al3 & $8i$ & \underline{1.028(8)} & 
0 & 0.25973(2) & 0.75139(3) & 0.00736(7) \\

Goodness of fit
& 1.062
& Al4 & $8h$ & 1.000(8) 
& 0.11374(2) & 0.11374(2) & $\frac{1}{2}$ & 0.00793(7) \\

$R_1$; $wR_2$ [$I>2\sigma(I)$]
& 0.0143; 0.0246 
& Al5 & $8j$ & 1.000(8)
& 0.82511(2) & 0.32511(2) & 0.48571(4) & 0.00941(7) \\

Diffraction peak/hole
(e$^{-}$/\AA$^3$)
& 0.353; $-0.248$ & \multicolumn{7}{c}{} \\


\hline
\hline

\multicolumn{9}{c}{
Step \#3
} \\

\hline
Space group; $Z$ 
& P4/nmm; 4 
& Atom 
& Wyckoff 
& Occupancy & $x$ & $y$ & $z$ 
& $U_{\rm eq}$ (\AA$^2$) \\

\cline{3-9}

$a$ (\AA)
& 12.8438(2)
& Mn1 & $8i$ & 0.997(6) 
& 0 & 0.24092(2) & 0.25446(2) & 0.00536(4) \\

$c$ (\AA)
& 5.1385(1)
& Ca1 & $2a$ & 1.004(6) 
& 0 & 0 & 0 & 0.00872(6) \\

Volume (\AA$^3$)
& 847.66(3)
& Ca2 & $2c$ & 1.004(6) 
& $\frac{1}{2}$ & 0 & 0.47507(3) & 0.00747(6) \\

No. of parameters
& 49
& Al1 & $8g$ & 1.002(6) 
& 0.17540(2) & 0.17540(2) & 0 & 0.00828(6) \\

$\theta$ range (deg)
& 1.769--29.167
& Al2 & $8j$ & 0.998(6)
& 0.89184(2) & 0.39184(2) & 0.03722(3) & 0.00849(6) \\

Refl. collected; independent refl.
& 13786; 1200
& Al3 & $8i$ & \underline{1.009(6)} 
& 0 & 0.25997(2) & 0.75139(2) & 0.007311(6) \\

Goodness of fit
& 1.166
& Al4 & $8h$ & 0.999(6) 
& 0.11376(2) & 0.11376(2) & $\frac{1}{2}$ & 0.00760(5) \\

$R_1$; $wR_2$ [$I>2\sigma(I)$]
& 0.0154; 0.0267 
& Al5 & $8j$ & 1.003(6) 
& 0.82501(2) & 0.32501(2) & 0.48551(3) & 0.00923(6) \\

Diffraction peak/hole
(e$^{-}$/\AA$^3$)
& 0.520; $-0.748$ & \multicolumn{7}{c}{} \\

\end{tabular}%
}
\end{ruledtabular}
\end{table*}
\renewcommand{\arraystretch}{1}

\begin{figure}[!t]
\centering 
\includegraphics[width=\columnwidth]{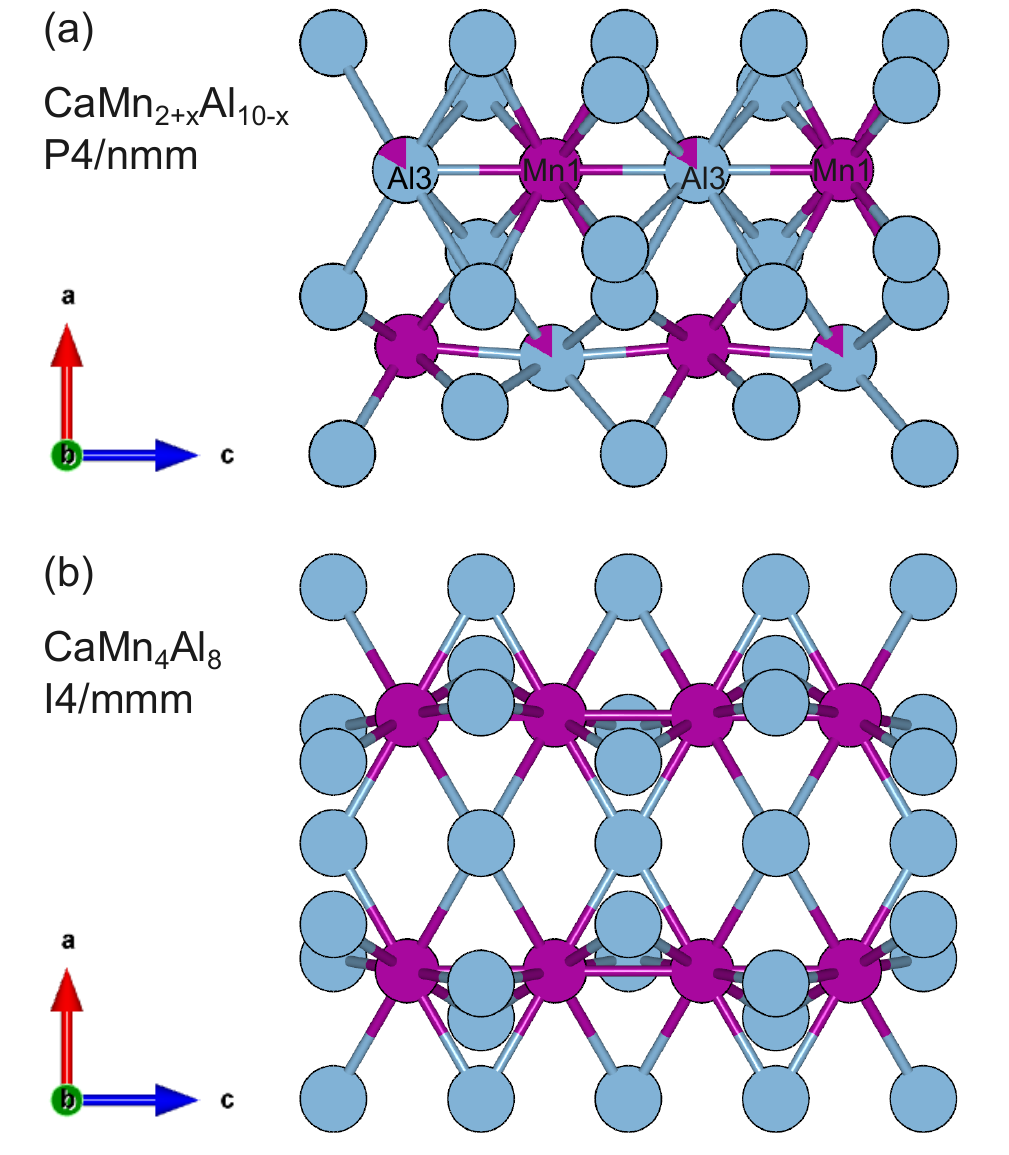}
\caption{ 
Crystal structures of (a) \ce{CaMn_{2+$x$}Al_{10-$x$}} and (b) \ce{CaMn4Al8}, emphasizing (a) Mn-Al-Mn, and (b)  Mn-Mn chains running along the $c$-axis. Mn atoms are shown in violet, Al in light blue, and Ca are omitted for clarity. In panel (a), mixed Mn/Al occupancy at the Al3 site is indicated.
}
\label{Fig:crystal-structure}
\end{figure}

We begin with an analysis of the crystal structure. \ce{CaMn_{2+$x$}Al_{10-$x$}} crystallizes in the \ce{CaCr2Al10}-type structure (\Cref{Fig:crystal-structure}a) \cite{Cordier1984, Thiede1998}, a ternary variant of the \ce{ThMn12}-type structure \cite{Florio1952}. Central to the discussion of its physical properties is the structural flexibility of ternary \ce{ThMn12}-type variants; in addition to 1-2-10 compounds \cite{Steinke2015, Sefat2007, Cordier1984, Thiede1998, Fulfer2012}, related 1-4-8 \cite{Fulfer2012, Felner1986} and 1-6-6 compounds \cite{Felner1982, Felner1986} are known. Many of these systems exhibit transition-metal/Al site mixing. Examples include \ce{RMn_{2+$x$}Al_{10-$x$}} (R = La, Gd, Yb), spanning compositions between 1-2-10 and 1-4-8 \cite{Sefat2007, Fulfer2012, Yanson1995}, and \ce{RFe_{4+$x$}Al_{8-$x$}} (R = Y, Lu, Gd, Er, Tb, Dy), between 1-4-8 and 1-6-6 \cite{Felner1986, Gorbunov2013, Waerenborgh2001, Liu1987, Felner1988, Srio2003}. Such site mixing can strongly influence the magnetic properties; for example, \ce{LaMn_{2+$x$}Al_{10-$x$}} was reported to exhibit spin-glass behavior \cite{Sefat2007}, whereas \ce{YFe_{4+$x$}Al_{8-$x$}} was reported to exhibit ferrimagnetism and spin-glass behavior \cite{Felner1986}. Therefore, for ternary \ce{ThMn12}-type aluminides, accounting for the homogeneity range and site mixing in the structural analysis is crucial for understanding their physical properties. A previous report found \ce{CaMn_{2+$x$}Al_{10-$x$}} to be stoichiometric ($x = 0$) \cite{Steinke2015}. To verify this result, SC-XRD measurements were performed on crystals obtained at each fractionation step. The SC-XRD data were initially refined using a model in which the occupancies of all atomic sites were allowed to vary. The results are summarized in \Cref{tab:SC-XRD-occ}. Within the 3$\sigma$ criterion, where $\sigma$ denotes the uncertainty, the occupancies of all sites except Al3 are equal to 1. For the Al3 site, an unphysical occupancy greater than 1 is statistically significant (>3$\sigma$) for the step \#1 and step \#2 crystals, whereas for step \#3, the deviation from full occupancy remains within 3$\sigma$.

\renewcommand{\arraystretch}{1.15}

\begin{table*}[ht]
\caption{Single-crystal X-ray diffraction refinement data, atomic positions,
site occupancies, and equivalent isotropic displacement parameters for
\ce{CaMn_{2+$x$}Al_{10-$x$}} crystals grown in steps \#1--3. Data were collected at room
temperature using Ag radiation ($\lambda = 0.56087$~\AA). $U_{\rm eq}$ is
defined as one-third of the trace of the orthogonalized $U_{ij}$ tensor
\cite{Fischer1988}. Mixed Mn/Al occupancy is included at the Al3 site.
}
\label{tab:SC-XRD-Mn}

\begin{ruledtabular}
\resizebox{\textwidth}{!}{%
\begin{tabular}{
    ll
    @{\hspace{1.1em}\vrule width 1.1pt\hspace{1.1em}}
    lcccccc
}


\multicolumn{9}{c}{
Step \#1, $x$ = 0.352(4)
} \\

\hline

Space group; $Z$
& P4/nmm; 4
& Atom & Wyckoff & Occupancy & $x$ & $y$ & $z$
& $U_{\rm eq}$ (\AA$^2$) \\

\cline{3-9}

$a$ (\AA)
& 12.8189(3)
& Mn1 & $8i$ & 1 
& 0 & 0.24236(2) & 0.25351(2) & 0.00651(4) \\

$c$ (\AA)
& 5.1445(2)
& Ca1 & $2a$ & 1 
& 0 & 0 & 0 & 0.00865(6) \\

Volume (\AA$^3$)
& 845.37(5)
& Ca2 & $2c$ & 1 
& $\frac{1}{2}$ & 0 & 0.47987(6) & 0.00749(8) \\

No. of parameters
& 42
& Al1 & $8g$ & 1 
& 0.17432(2) & 0.17432(2) & 0 & 0.00974(6) \\

$\theta$ range (deg)
& 1.773--31.800
& Al2 & $8j$ & 1 
& 0.89051(2) & 0.39051(2) & 0.03148(5) & 0.00950(6) \\

Refl. collected; independent refl.
& 23002; 1554
& Al3 & $8i$ & 0.824(2) 
& 0 & 0.25832(2) & 0.75103(4) & 0.00749(8) \\

Goodness of fit
& 1.027
& Mn2 & $8i$ & 0.176(2) 
& 0 & 0.25832(2) & 0.75103(4) & 0.00749(8) \\

$R_1$; $wR_2$ [$I>2\sigma(I)$]
& 0.0290; 0.0345
& Al4 & $8h$ & 1 
& 0.11344(2) & 0.11344(2) & $\frac{1}{2}$ & 0.00900(6) \\

Diffraction peak/hole
(e$^{-}$/\AA$^3$)
& 0.569; $-0.468$
& Al5 & $8j$ & 1 
& 0.82592(2) & 0.32592(2) & 0.48734(5) & 0.01027(6) \\


\hline
\hline

\multicolumn{9}{c}{
Step \#2, $x$ = 0.062(4)
} \\

\hline

Space group; $Z$
& P4/nmm; 4
& Atom & Wyckoff & Occupancy & $x$ & $y$ & $z$
& $U_{\rm eq}$ (\AA$^2$) \\

\cline{3-9}

$a$ (\AA)
& 12.8664(4)
& Mn1 & $8i$ & 1 
& 0 & 0.24109(2) & 0.25436(2) & 0.00566(3) \\

$c$ (\AA)
& 5.1453(2)
& Ca1 & $2a$ & 1 
& 0 & 0 & 0 & 0.00870(5) \\

Volume (\AA$^3$)
& 851.77(6)
& Ca2 & $2c$ & 1 
& $\frac{1}{2}$ & 0 & 0.47558(5) & 0.00728(5) \\

No. of parameters
& 42
& Al1 & $8g$ & 1 
& 0.17529(2) & 0.17529(2) & 0 & 0.00851(5) \\

$\theta$ range (deg)
& 1.766--29.209
& Al2 & $8j$ & 1 
& 0.89166(2) & 0.39166(2) & 0.03659(4) & 0.00875(5) \\

Refl. collected; independent refl.
& 12001; 1181
& Al3 & $8i$ & 0.969(2) 
& 0 & 0.25974(2) & 0.75138(3) & 0.00751(8) \\

Goodness of fit
& 1.065
& Mn2 & $8i$ & 0.031(2) 
& 0 & 0.25974(2) & 0.75138(3) & 0.00751(8) \\

$R_1$; $wR_2$ [$I>2\sigma(I)$]
& 0.0143; 0.0246
& Al4 & $8h$ & 1 
& 0.11374(2) & 0.11374(2) & $\frac{1}{2}$ & 0.00792(5) \\

Diffraction peak/hole
(e$^{-}$/\AA$^3$)
& 0.346; $-0.273$ 
& Al5 & $8j$ & 1 
& 0.82511(2) & 0.32511(2) & 0.48571(4) & 0.00939(5) \\


\hline
\hline

\multicolumn{9}{c}{
Step \#3, $x$ = 0.024(4)
} \\

\hline
Space group; $Z$ 
& P4/nmm; 4 
& Atom 
& Wyckoff 
& Occupancy & $x$ & $y$ & $z$ 
& $U_{\rm eq}$ (\AA$^2$) \\

\cline{3-9}

$a$ (\AA)
& 12.8438(2)
& Mn1 & $8i$ & 1 
& 0 & 0.24092(2) & 0.25446(2) & 0.00542(4) \\

$c$ (\AA)
& 5.1385(1)
& Ca1 & $2a$ & 1 
& 0 & 0 & 0 & 0.00859(4) \\

Volume (\AA$^3$)
& 847.66(3)
& Ca2 & $2c$ & 1 
& $\frac{1}{2}$ & 0 & 0.47506(4) & 0.00733(4) \\

No. of parameters
& 42
& Al1 & $8g$ & 1 
& 0.17540(2) & 0.17540(2) & 0 & 0.00819(4) \\

$\theta$ range (deg)
& 1.769--29.167
& Al2 & $8j$ & 1 
& 0.89184(2) & 0.39184(2) & 0.03722(3) & 0.00853(4) \\

Refl. collected; independent refl.
& 13786; 1200
& Al3 & $8i$ & 0.988(2) 
& 0 & 0.25997(2) & 0.75139(2) & 0.00741(7) \\

Goodness of fit
& 1.166
& Mn2 & $8i$ & 0.012(2)
& 0 & 0.25997(2) & 0.75139(2) & 0.00741(7) \\

$R_1$; $wR_2$ [$I>2\sigma(I)$]
& 0.0155; 0.0271
& Al4 & $8h$ & 1 
& 0.11377(2) & 0.11377(2) & $\frac{1}{2}$ & 0.00762(4) \\

Diffraction peak/hole
(e$^{-}$/\AA$^3$)
& 0.520; $-0.753$
& Al5 & $8j$ & 1 
& 0.82501(2) & 0.32501(2) & 0.48550(3) & 0.00912(4) \\

\end{tabular}%
}
\end{ruledtabular}
\end{table*}
\renewcommand{\arraystretch}{1}

An occupancy greater than 1 indicates that the stoichiometric model underestimates the electron density at this site. We therefore tested the possibility of Mn/Al site mixing at Al3 site, since Mn is electron-rich compared to Al, and such mixing is known to occur in structurally-related 1-2-10 compounds \cite{Fulfer2012, Sefat2007}. \Cref{tab:SC-XRD-Mn} presents refinements of the same raw data using a model that allows mixed Mn/Al occupancy at the Al3 site. All other site occupancies were fixed at 1, since refining them as free parameters yielded full occupancies within $3\sigma$ (\Cref{tab:SC-XRD-occ}). The quality of the refinements allowing for the Mn/Al site mixing, as judged by the $R_1$ and $wR_2$ values, is comparable to that obtained with freely refined occupancies, while avoiding the unphysical occupancy greater than 1. The mixed-site model is statistically justified for step \#1 and step \#2 crystals, but the Al3 occupancy for step \#3 is equal to 1 within $3\sigma$. Therefore, the result for step \#3 should be interpreted with caution. Nevertheless, magnetization and transport measurements presented later in this work are consistent with the presence of a small amount of excess Mn at each fractionation step, including step \#3 crystals.

The stoichiometries determined by EDS (\Cref{Fig:EDS-ratios} in the Supporting Information) and SC-XRD (\Cref{tab:SC-XRD-occ} and \Cref{tab:SC-XRD-Mn}) are summarized in \Cref{tab:stoichiometries}. Within $3\sigma$, the stoichiometries obtained from EDS are indistinguishable. The EDS results are consistent with the \ce{CaMn_{2+$x$}Al_{10-$x$}} phase at each step; however, without in-situ standardization, they are not sufficiently sensitive to resolve the small differences in excess Mn concentration between consecutive fractionation steps. In contrast, refinement of the SC-XRD data indicates a substantially larger excess Mn occupancy on the Al3 site in step \#1 crystals, whereas step \#2 and step \#3 crystals contain small but still detectable amounts of excess Mn, suggesting a finite width of formation in \ce{CaMn_{2+$x$}Al_{10-$x$}}.

The origin of the preference for Mn/Al mixing at the Al3 site is illustrated in \Cref{Fig:crystal-structure}. A structurally-related compound, \ce{CaMn4Al8}, contains linear Mn--Mn chains extending along the $c$-axis. In \ce{CaMn_{2+$x$}Al_{10-$x$}}, the different space-group symmetry instead gives rise to corrugated Mn--Al--Mn chains, with Al3 being the site that connects two Mn atoms. Among the Al sites in the structure, it is therefore the most plausible site for Mn substitution; full replacement of Al by Mn at this position would generate a local motif characteristic of the stable \ce{CaMn4Al8} phase. The same behavior was found to occur in \ce{RMn_{2+$x$}Al_{10-$x$}} (R = La, Gd), where $8i$ Al position was found to be partially occupied by Mn \cite{Sefat2007}.


\begin{table}[h]
\caption{Stoichiometries of the \ce{CaMn_{2+$x$}Al_{10-$x$}} crystals grown in steps
\#1--3, as determined by energy-dispersive X-ray spectroscopy (EDS) and
single-crystal X-ray diffraction (SC-XRD).}
\label{tab:stoichiometries}
\begin{ruledtabular}
\begin{tabular}{ccc}
Step \# &
EDS &
SC-XRD \\
\hline
1 & 
\ce{CaMn_{2.34(14)}Al_{10.09(34)}} & 
\ce{CaMn_{2.352(4)}Al_{9.648(4)}}  \\

2 & 
\ce{CaMn_{2.27(6)}Al_{10.19(11)}} & 
\ce{CaMn_{2.062(4)}Al_{9.938(4)}}  \\

3 & 
\ce{CaMn_{2.26(4)}Al_{9.89(24)}} & 
\ce{CaMn_{2.024(4)}Al_{9.976(4)}}  \\
\end{tabular}
\end{ruledtabular}
\end{table}


\subsection{Magnetization}

\begin{figure}[ht!]
\centering 
\includegraphics[width=0.4\textwidth]{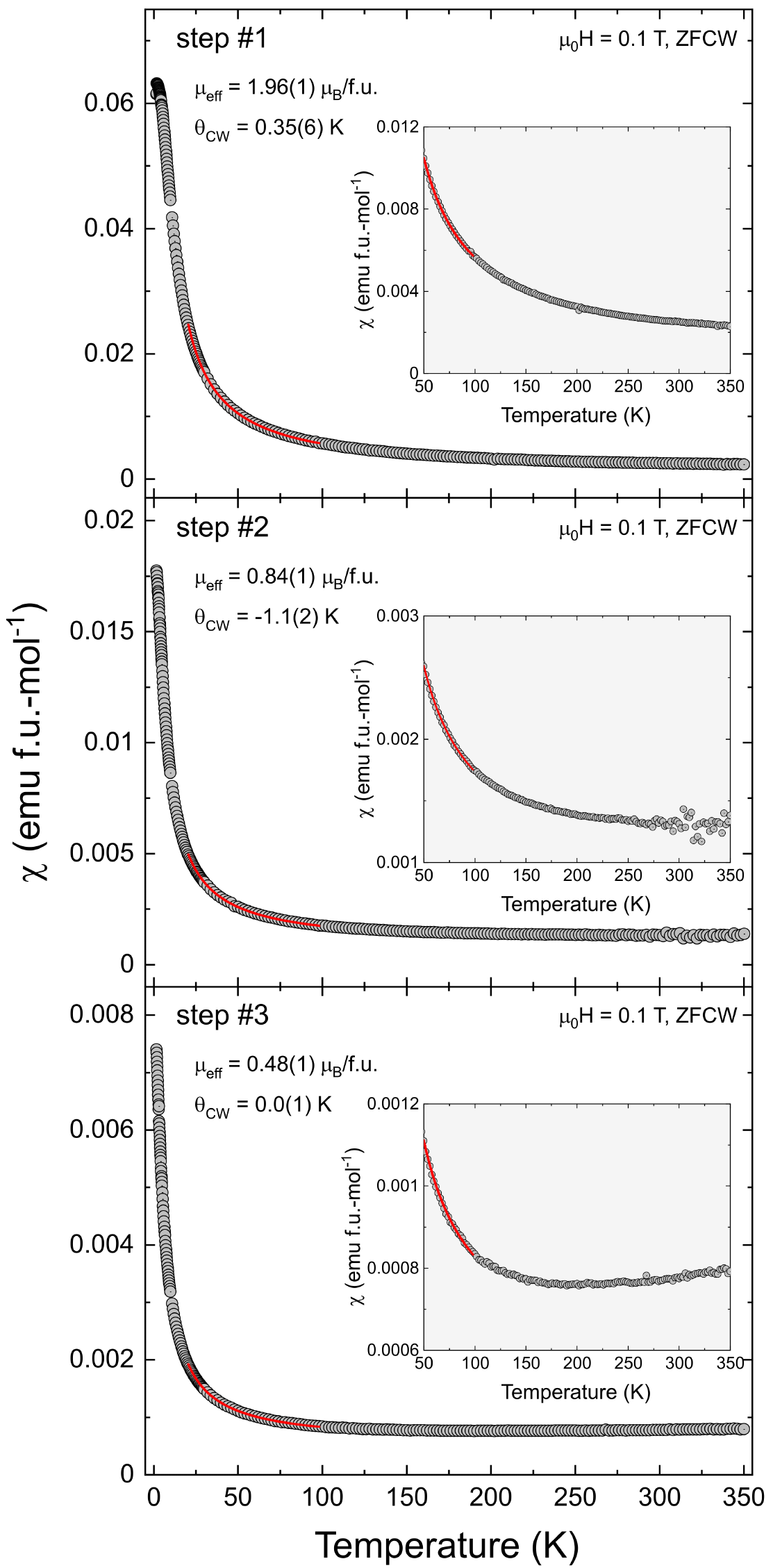}
\caption{ 
Temperature-dependent magnetic susceptibility of \ce{CaMn_{2+$x$}Al_{10-$x$}} single crystals measured with magnetic field parallel to the $c$-axis. The low-temperature upturn was modeled using a Curie-Weiss law (red) in the 20 – 100 K temperature range (see text for details). The insets show a zoomed-in view of the susceptibility, emphasizing non-Curie-Weiss behavior at high temperatures, most clearly observed for the step \#2 and \#3 crystals.
}
\label{Fig:MPMS-CW}
\end{figure}

\begin{figure}[htbp]
\centering 
\includegraphics[width=\columnwidth]{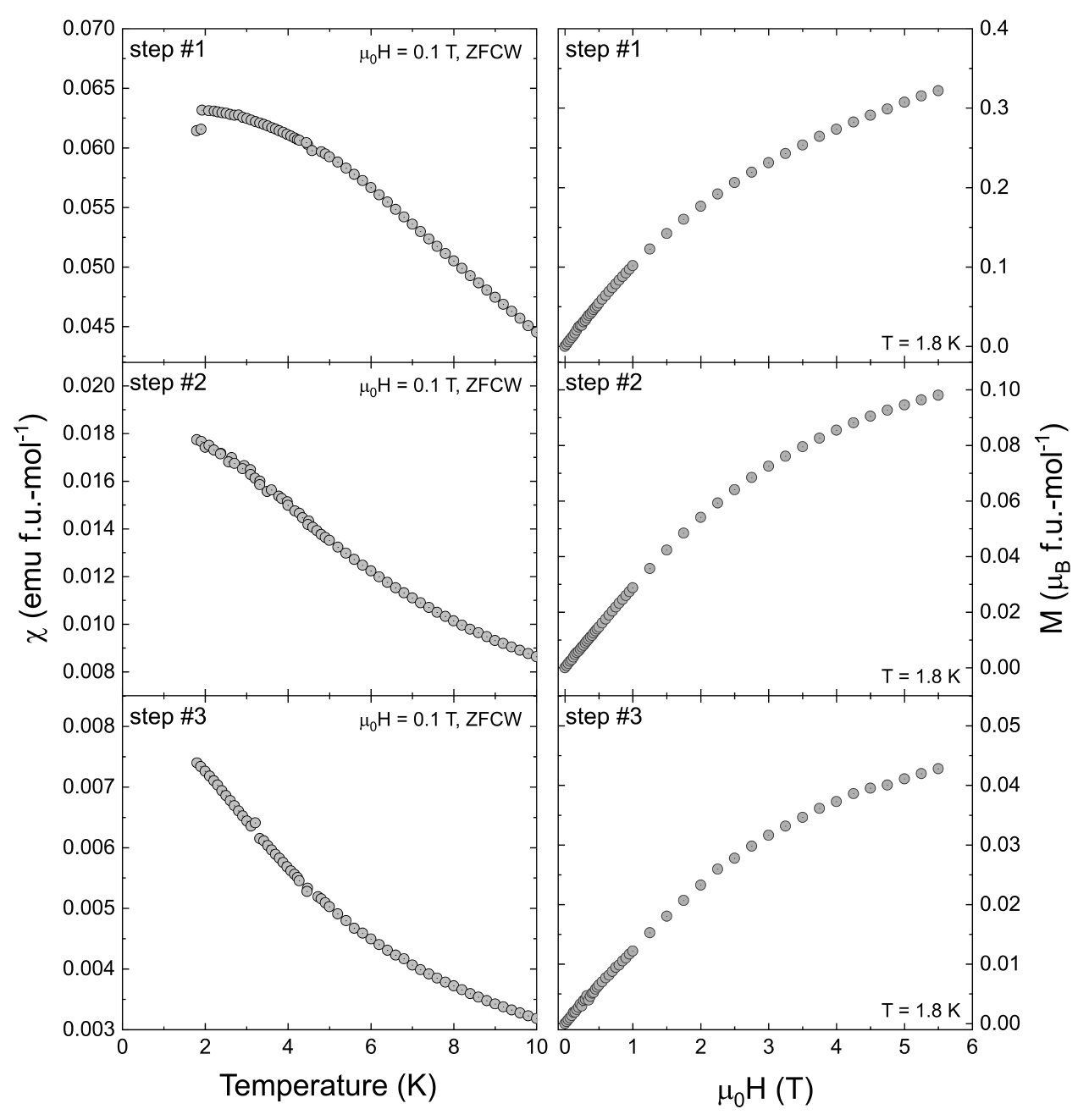}
\caption{ 
Low-temperature temperature-dependent magnetic susceptibility ($\chi(T)$; left panel), and field-dependent magnetization ($M(H)$; right panel) of \ce{CaMn_{2+$x$}Al_{10-$x$}} single crystals, measured with magnetic field parallel to the $c$-axis. The saturation of $\chi(T)$ becomes progressively weaker between steps \#1 and \#3, accompanied by a corresponding decrease in $M(H)$.
}
\label{Fig:MPMS-saturation}
\end{figure}

To determine how Mn/Al site mixing affects the physical properties, temperature-dependent magnetization of \ce{CaMn_{2+$x$}Al_{10-$x$}} single crystals was measured. \Cref{Fig:MPMS-CW} shows the temperature-dependent magnetic susceptibility together with Curie--Weiss fits over the temperature range 20--100 K (red lines). The same fit range was used for each step to ensure consistency; see \Cref{Fig:CW-sanity} in the Supporting Information for details. The results of the Curie--Weiss analysis are summarized in \Cref{tab:impurity_moments}. The Curie--Weiss temperatures are close to zero for the three fractionation steps, indicating negligible magnetic interactions between Mn atoms. The effective moment per excess Mn, $\mu_{\mathrm{eff/Mn}}$, was calculated as $\mu_{\mathrm{eff/Mn}} = \mu_{\mathrm{eff/f.u.}}/\sqrt{x}$, using the relations $C_{\mathrm{f.u.}} = xC_{\mathrm{Mn}}$ and $\mu_{\mathrm{eff}} = \sqrt{8C}$, where $C$ is the Curie constant. Magnetization at $T$ = 1.8 K and $\mu_{\mathrm{0}}$H = 5.5 T per excess Mn, $M_{\mathrm{Mn}}$(1.8 K; 5.5 T), was calculated as $M_{\mathrm{f.u.}}$(1.8 K; 5.5 T)/$x$. Within the experimental uncertainty, the effective moment per excess Mn is similar for all three fractionation steps, with $\mu_{\mathrm{eff/Mn}} \approx 3\ \mu_{\mathrm{B}}/\mathrm{Mn}$. This consistency supports the interpretation that the magnetism in \ce{CaMn_{2+$x$}Al_{10-$x$}} originates from excess Mn atoms occupying the Al3 site. Although this value does not correspond to the spin-only moment of free \ce{Mn^{2+}} ($5.93\ \mathrm{\mu_B}$, high spin) or \ce{Mn^{3+}} ($4.90\ \mathrm{\mu_B}$, high spin) ion, excess Mn atoms in \ce{CaMn_{2+$x$}Al_{10-$x$}} hybridize with the surrounding Al atoms, which can result in the reduced value of the effective moment. In fact, if the stoichiometric \ce{CaMn2Al10} has $\mathrm{\mu_{eff/f.u.} = 0}$, as the data suggest, then Mn--Al hybridization could be strong enough to fully suppress the fluctuating Mn moment for $x = 0$. The values of $M_{\mathrm{Mn}}$(1.8 K; 5.5 T), as determined from \Cref{Fig:MPMS-saturation}, are identical for steps \#2 and \#3 within experimental uncertainty, whereas the value for step \#1 is substantially smaller. This behavior may be related to stronger Mn--Mn interactions in step \#1 crystal, which has almost six times the concentration of excess Mn atoms as step \#2 crystal. At higher concentrations, the relatively stronger interactions between magnetic atoms can inhibit the ability of impurity spins to align with magnetic field \cite{Smith}.

Further support for the interpretation that the magnetic properties of \ce{CaMn_{2+$x$}Al_{10-$x$}} originate from excess Mn atoms is provided by the low-temperature magnetic susceptibility, $\chi(T)$, and field-dependent magnetization, $M(H)$, as shown in \Cref{Fig:MPMS-saturation}. The magnetic susceptibilities of \ce{CaMn_{2+$x$}Al_{10-$x$}} crystals from all three fractionation steps show no evidence of a magnetic transition. As the excess Mn concentration decreases from step \#1 to step \#3, the tendency of $\mathrm{\chi(T)}$ to saturate weakens. At the same time, the value of $\mathrm{M(H)}$ at T = 1.8 K and $\mathrm{\mu_0H} = 5.5$ T decreases. Both trends are consistent with a decreasing concentration of excess Mn, as observed in SC-XRD measurements. For step \#1, $\mathrm{\chi(T)}$ shows a step-like decrease below T $\approx$ 2 K, which may indicate the onset of spin freezing, although AC susceptibility measurements below $T$ = 1.8 K would be better suited to confirm this interpretation. A previous study \cite{Steinke2015}, which reported \ce{CaMn_{2+$x$}Al_{10-$x$}} to be stoichiometric, i.e., \ce{CaMn2Al10}, likewise found no magnetic transition over the temperature range 1.8--400 K. The reported magnetic susceptibility deviated from Curie--Weiss behavior above approximately 250 K, although this deviation was not discussed by the authors. The absence of magnetic ordering, together with the low-temperature AC susceptibility following $\chi'(T) \sim T^{-1.2}$, was interpreted as evidence for incipient ferromagnetism with a Curie temperature below $T = 2$ K \cite{Steinke2015}. The reported effective moment was $0.83\ \mu_{\mathrm{B}}/\mathrm{Mn}$, determined from the DC susceptibility normalized per Mn atom. The authors of Ref. \cite{Steinke2015} assumed $x$ = 0 (\ce{CaMn2Al10}), thus the effective moment per formula unit can be calculated as $\mu_{eff/f.u.} = \sqrt{2}\mu_{eff/Mn}$, yielding $1.17\ \mu_{\mathrm{B}}/\mathrm{f.u.}$ This value lies in between 
those of the step \#1 ($\mu_{\mathrm{eff}} = 1.96(1)\ \mu_{\mathrm{B}}/\mathrm{f.u.}$) and step \#2 ($\mu_{\mathrm{eff}} = 0.84(1)\ \mu_{\mathrm{B}}/\mathrm{f.u.}$) crystals. It is therefore possible that the samples studied in \cite{Steinke2015} contained some excess Mn. Unfortunately, as mentioned above in the growth section, although the physical properties of \ce{CaMn2Al10} were reported, the growth profile was not, making an exact reproduction of the growth conditions impossible. Moreover, the crystallographic site occupancies were not included, and the full occupancy and stoichiometric nature of \ce{CaMn2Al10} were inferred only from the reasonable values of the $R_1$ and $wR_2$ indices and the ease with which the structure was solved using the charge-flipping algorithm. This is a serious concern for ternary \ce{CaCr2Al10}-type aluminides because, as discussed in the crystal structure section, these compounds often exhibit transition-metal/Al site mixing, which needs to be carefully verified. Moreover, a small amount of excess Mn may produce only minor changes in the refinement indices and may even remain below the detection limit of laboratory SC-XRD. At the same time, paramagnetic impurities at concentrations below 1\% can readily produce detectable signatures in magnetization \cite{Loram1970} and/or electrical resistivity \cite{Ford1970} measurements. Taking the above into account, the magnetic properties previously attributed to stoichiometric \ce{CaMn2Al10} may instead originate from a small amount of excess Mn in \ce{CaMn_{2+$x$}Al_{10-$x$}} arising from Mn/Al site mixing.

\begin{table}[t]
\caption{Effective moments, magnetization at $T$ = 1.8 K and $\mu_{\mathrm{0}}$H = 5.5 T, and Curie-Weiss temperatures of the \ce{CaMn_{2+$x$}Al_{10-$x$}} crystals grown in steps \#1--3, with excess Mn concentration, $x$, determined from SC-XRD analysis. Values per f.u. are obtained directly from Curie–Weiss fits (\Cref{Fig:MPMS-CW}); see text for details of the calculation of $\mu_{\mathrm{eff/Mn}}$ and $M_{\mathrm{Mn}}$(1.8 K; 5.5 T).
}

\label{tab:impurity_moments}

\begin{ruledtabular}
\begin{tabular}{cccccc}

Step &
$x$ &
\makecell{$\theta_{\mathrm{CW}}$ \\ (K)} &
\makecell{$\mu_{\mathrm{eff/f.u.}}$ \\ ($\mu_{\mathrm{B}}$)} &
\makecell{$\mu_{\mathrm{eff/Mn}}$ \\ ($\mu_{\mathrm{B}}$)} &
\makecell{$M_{\mathrm{Mn}}$(1.8 K; 5.5 T) \\ ($\mu_{\mathrm{B}}$)} \\
\hline

\#1 & 0.352(4) & 0.35(6) & 1.96(1)  &  3.30(3)   & 0.92(1)  \\
\#2 & 0.062(4) & -1.1(2) & 0.84(1)  &  3.37(12)  & 1.58(10)  \\
\#3 & 0.024(4) & 0.0(1)  & 0.48(1)  &  3.10(27)  & 1.79(30)  \\

\end{tabular}
\end{ruledtabular}
\end{table}

\subsection{Electrical resistivity}

\begin{figure}[htbp]
\centering 
\includegraphics[width=\columnwidth]{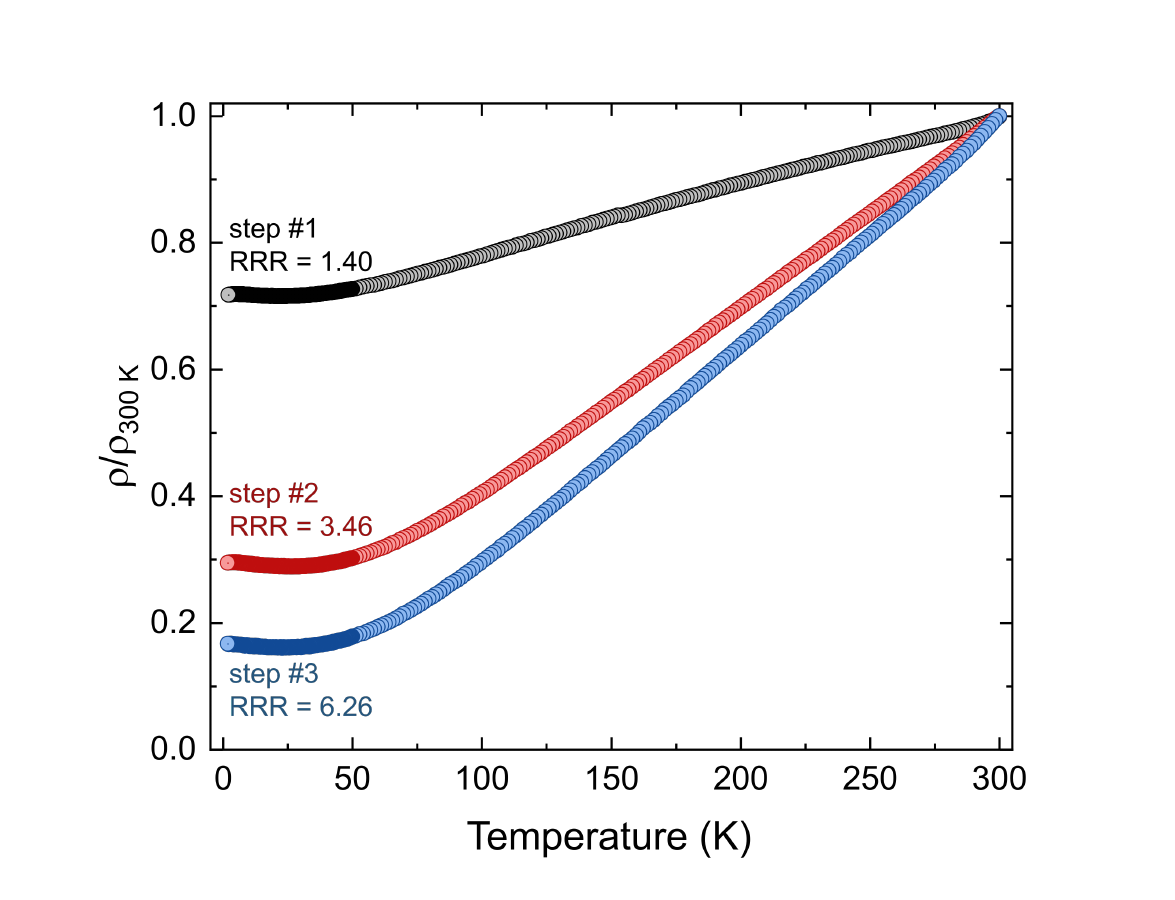}
\caption{
Temperature-dependent zero-field electrical resistivity of \ce{CaMn_{2+$x$}Al_{10-$x$}} single crystals. The single crystals exhibit progressively larger RRR values from step \#1 to step \#3, consistent with a reduction of the degree of site disorder, x, associated with excess Mn.
}
\label{Fig:rho-RRR}
\end{figure}

\begin{figure}[htbp]
\centering 
\includegraphics[width=0.35\textwidth]{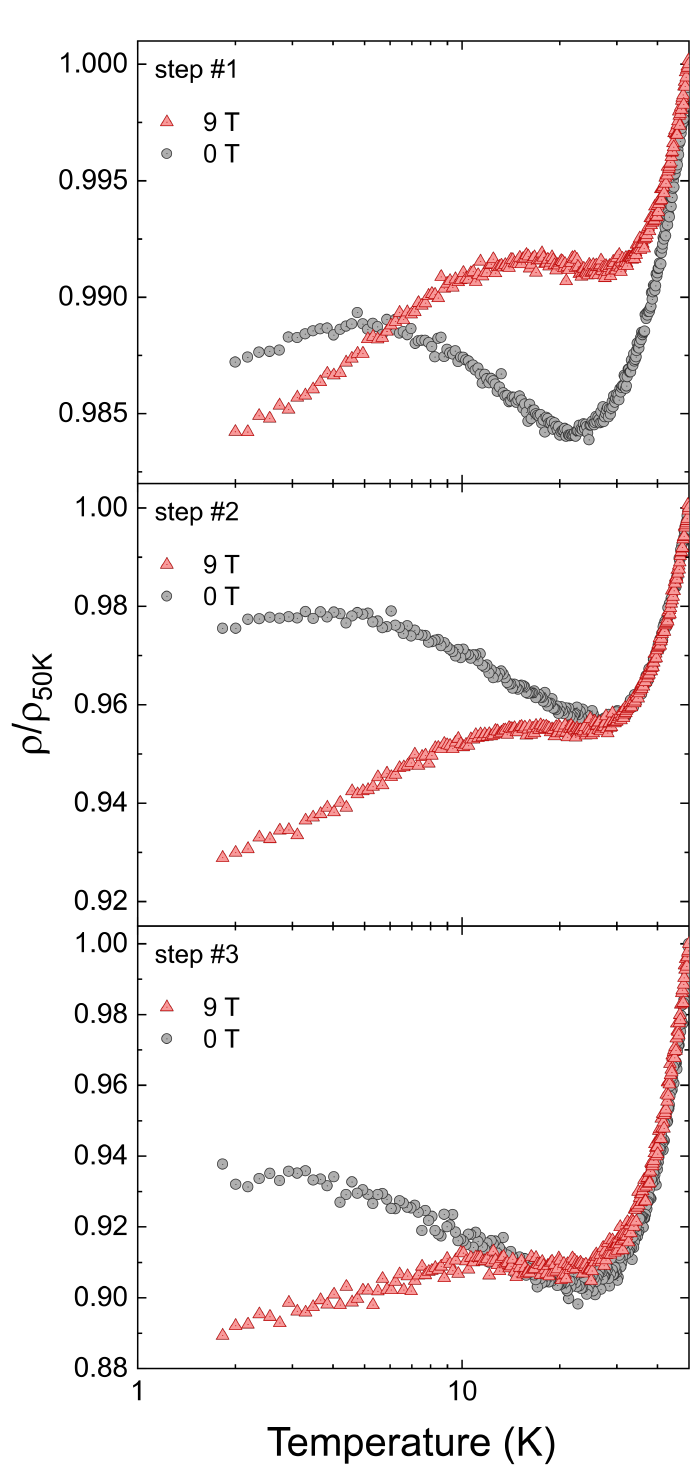}
\caption{ 
Temperature-dependent normalized electrical resistivity, $\mathrm{\rho/\rho_{50K}}$, of \ce{CaMn_{2+$x$}Al_{10-$x$}} single crystals, measured with magnetic field perpendicular to the $c$-axis. The field-dependent resistivity maximum gets progressively weaker between steps \#1 and \#3.
}
\label{Fig:rho-low-temps}
\end{figure}

The effect of Mn/Al site mixing on the electronic transport was further examined with temperature-dependent electrical resistivity measurements, shown in \Cref{Fig:rho-RRR}. As the excess Mn concentration, x, decreases, the residual resistivity ratio (RRR), defined here as $\rho_{300\mathrm{K}}/\rho_\mathrm{0}$, where $\rho_\mathrm{0}$ is the resistivity at the minimum (see \Cref{Fig:rho-low-temps}), increases from 1.40 for step \#1 to 6.26 for step \#3, indicating substantially reduced scattering across the fractionation sequence. The RRR of the sample measured in Ref.~\cite{Steinke2015} was not reported by the authors, but can be estimated from the published data as $RRR \approx  100 \ \mu \Omega \ \mathrm{cm} /30 \ \mu \Omega \ \mathrm{cm}        \ \approx 3$. This value lies between those obtained for the step \#1 and step \#2 crystals, as does the effective moment per formula unit determined from magnetic susceptibility measurements, providing further support for the argument that the \ce{CaMn2Al10} single crystals studied in Ref. \cite{Steinke2015} contained a small amount of excess Mn arising from Mn/Al site mixing.

The low-temperature resistivity, plotted on a semilogarithmic temperature scale, is presented in \Cref{Fig:rho-low-temps}. A low-temperature upturn is observed, which in a system without long-range magnetic order may arise from several mechanisms: the Kondo effect \cite{Kondo1964, Krlak2025}, spin-glass freezing \cite{Schilling1976, Sefat2007}, or quantum corrections to the resistivity, including Altshuler--Aronov \cite{Altshuler1993, Gnida2021} and weak-localization effects \cite{Abrahams1979, Bergmann1984}. The feature observed in \ce{CaMn_{2+$x$}Al_{10-$x$}} is field-dependent, which excludes the Altshuler--Aronov effect as the dominant contribution. Moreover, the presence of a resistivity maximum in the step \#1 sample, and to a lesser extent in the step \#2 sample, strongly suggests spin-glass-like freezing. Weak localization is not expected to produce such a maximum, and in the single-ion Kondo regime the resistivity is expected to saturate as the unitarity limit is approached \cite{vanderWiel2000}. For spin-glass freezing, there is a competition between the single-ion Kondo effect and interactions between magnetic atoms. As the concentration of magnetic atoms increases, these interactions increasingly “lock in” the moment directions, weakening the Kondo resonance and giving rise to a maximum in the resistivity \cite{Schilling1976}. Thus, for the step \#1 crystal, which has the largest concentration of excess Mn, interactions between Mn atoms are the strongest and the resistivity maximum is clearly visible. As the concentration of excess Mn decreases, the relative importance of the single-ion Kondo effect increases: the step \#2 crystals exhibit only a weak maximum, whereas the step \#3 crystals show behavior closer to saturation.

In a previous report, the maximum in electrical resistivity, together with maxima in AC susceptibility and specific heat were suggested to signal "\textit{a low-lying energy scale where a gap opens for the critical fluctuations associated with this incipient magnetic order}" \cite{Steinke2015}. However, the main evidence supporting this interpretation was the scaling behavior of the AC susceptibility, while non-Fermi-liquid behavior was observed neither in the specific heat nor in the resistivity. In view of this, together with the magnetization and resistivity results presented here, we find the interpretation in terms of intrinsic incipient ferromagnetism to be insufficiently supported. A more plausible scenario is that the observed behavior originates from excess Mn associated with the finite width of formation in \ce{CaMn_{2+$x$}Al_{10-$x$}}.

\section{CONCLUSIONS}

In this work, we demonstrated the controlled growth of three batches of \ce{CaMn_{2+$x$}Al_{10-$x$}} single crystals using a fractionation approach. Single-crystal X-ray diffraction measurements revealed that the crystals were nonstoichiometric, $x \neq 0$. The influence of excess Mn was investigated through magnetization and electrical resistivity measurements, which revealed spin-glass-like features, including a low-temperature upturn in the magnetic susceptibility and a maximum in the electrical resistivity. The properties of \ce{CaMn_{2+$x$}Al_{10-$x$}} were discussed in relation to the previously reported incipient ferromagnetism in stoichiometric \ce{CaMn2Al10}, with the present work instead indicating that the observed magnetic behavior is more likely associated with Mn/Al site mixing. This work demonstrates the usefulness of fractionation as a crystal-growth strategy for itinerant magnetic systems, in which the magnetic response can be intrinsically weak and therefore easily obscured by extrinsic effects, such as small deviations from stoichiometry. By providing several compositions grown from the same initial charge, fractionation enables an unambiguous analysis of structure–property relationships.

\section{ACKNOWLEDGEMENTS}

This work was supported by the U.S. Department of Energy, Office of Basic Energy Science, Division of Materials Sciences and Engineering. The research was performed at the Ames National Laboratory. Ames National Laboratory is operated for the U.S. Department of Energy by Iowa State University under Contract No. DE-AC02-07CH11358. Szymon Kr{\'o}lak acknowledges support from the Polish National Agency for Academic Exchange (NAWA) under the Bekker Programme (no. BPN/BEK/2024/1/00041). 

Szymon Kr{\'o}lak used GPT-5.6 Sol for grammar correction, and the generated text was carefully checked afterwards.

\section{DATA AVAILABILITY}
The data that support the findings of this study will be made available upon acceptance of the manuscript.

\nocite{*}
\bibliography{reference}


\clearpage
\onecolumngrid

\begin{center}
{\large\bfseries Supporting Information}

\vspace{1.5em}

{\large\bfseries
Fractionation as a Tool to Control Mn/Al Site Mixing and Magnetism in \ce{CaMn_{2+$x$}Al_{10-$x$}}
}

\vspace{1em}

S. Kr\'{o}lak,$^{1,2,3}$
S. Kumari,$^{1,2}$
S. L. Bud’ko,$^{1,2}$
P. C. Canfield,$^{1,2}$

\vspace{0.7em}

{\itshape
$^{1}$Ames National Laboratory, U.S. Department of Energy, Iowa State University, Ames, Iowa 50011, USA\\
$^{2}$Department of Physics and Astronomy, Iowa State University, Ames, Iowa 50011, USA\\
$^{3}$Faculty of Applied Physics and Mathematics and Advanced Materials Center, Gdansk University of Technology, Narutowicza 11/12, 80-233 Gdansk, Poland\\
}
\end{center}

\vspace{2em}

\twocolumngrid

\setcounter{figure}{0}
\setcounter{table}{0}
\renewcommand{\thefigure}{S\arabic{figure}}
\renewcommand{\thetable}{S\arabic{table}}

\begin{figure}[htbp]
\centering 
\includegraphics[width=\columnwidth]{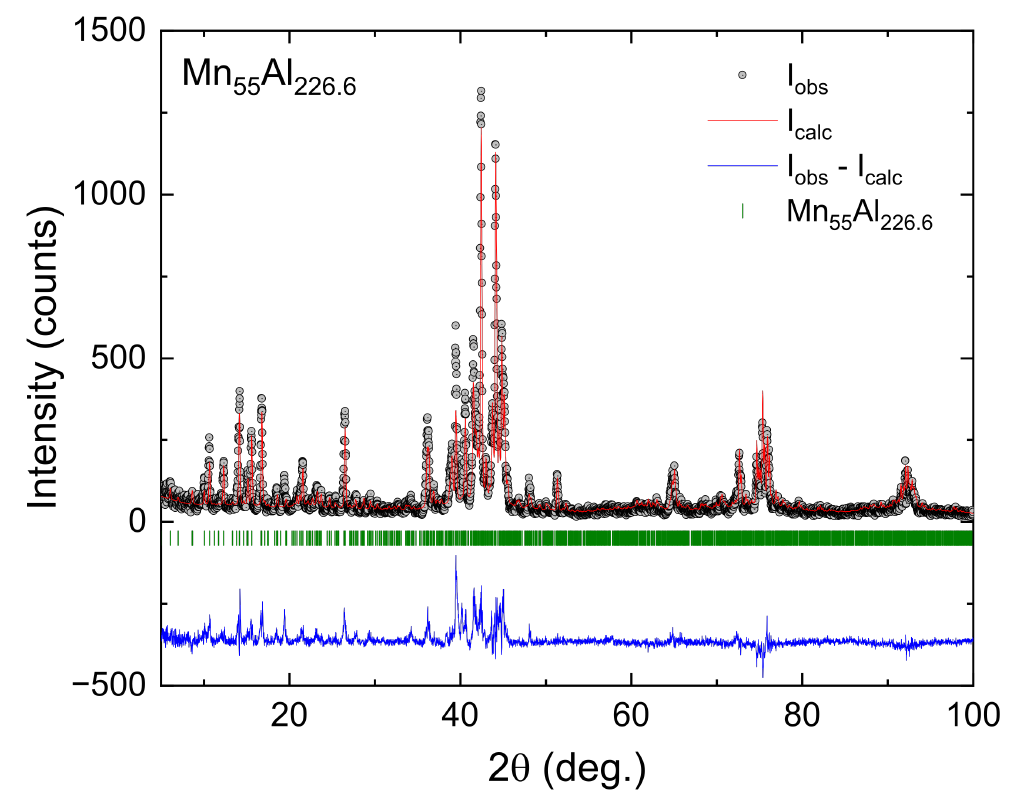}
\caption{ 
Powder X-ray diffraction pattern of \ce{Mn_{55}Al_{226.6}}. Experimental data are shown as black circles, along the Rietveld refinement (red), difference between the measured data and the model (blue), and the Bragg reflection positions (green).
}
\label{Fig:Mn55_pXRD}
\end{figure}

\Cref{Fig:Mn55_pXRD} shows the powder X-ray diffraction pattern of ground \ce{Mn_{55}Al_{226.6}} single crystals. Owing to the complexity of the crystal structure, which contains 42 atoms per unit cell, the atomic positions, site occupancies, and thermal displacement parameters were not refined and were instead fixed to the values determined from single-crystal X-ray diffraction [1]. The refined lattice constants are $a=b=20.006(2)$ \AA\ and $c=24.732(1)$ \AA, in good agreement with the literature values [1].

\begin{figure}[ht!]
\centering 
\includegraphics[width=1.1\columnwidth]{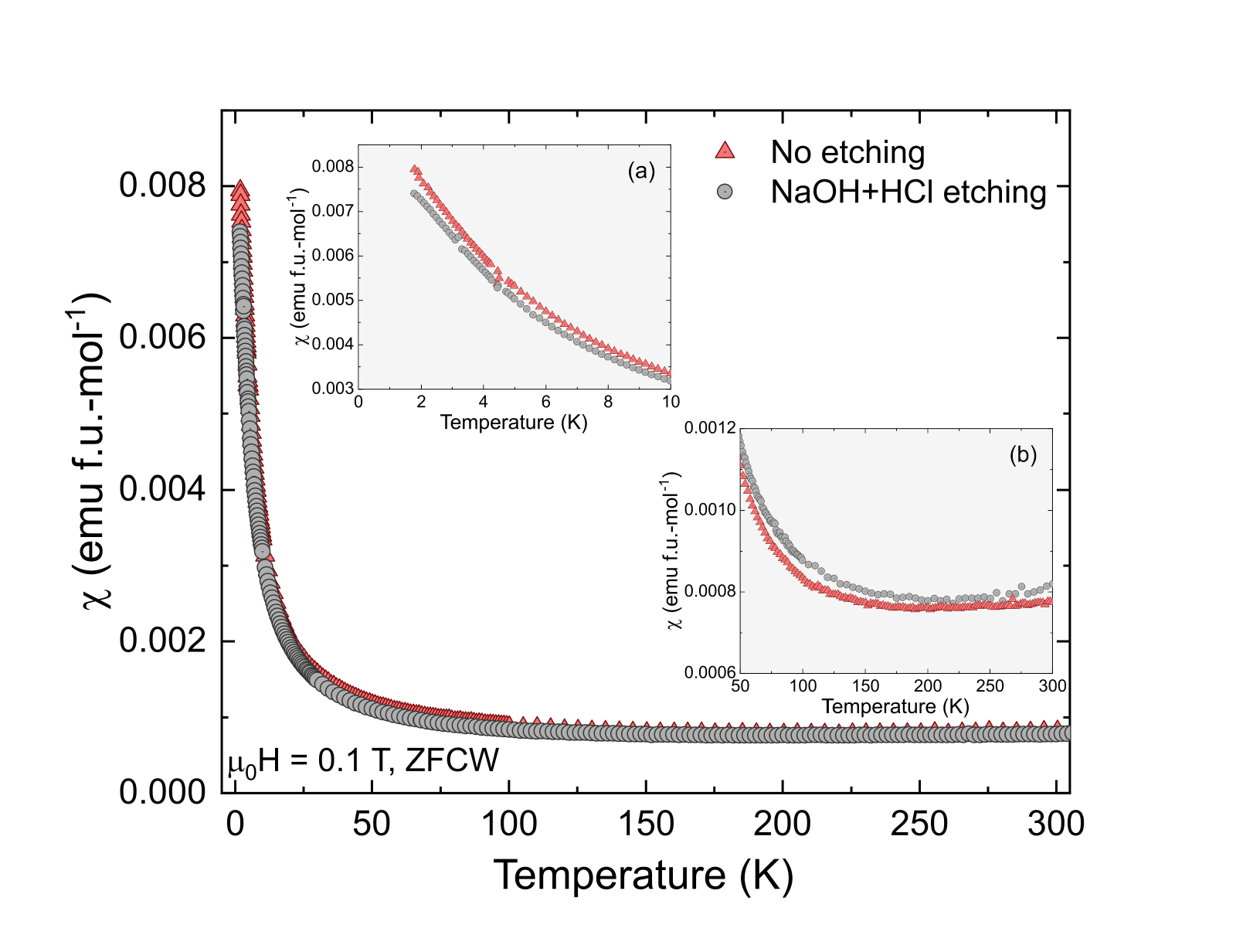}
\caption{ 
Temperature-dependent magnetic susceptibility of \ce{CaMn_{2+$x$}Al_{10-$x$}} single crystals grown in step \#3, measured with the magnetic field applied parallel to the crystallographic $c$-axis before and after chemical etching. Insets (a) and (b) show the low- and high-temperature regions, respectively.
}
\label{Fig:etching}
\end{figure}

As-grown single crystals of \ce{CaMn_{2+$x$}Al_{10-$x$}} were covered with residual Al flux. To prepare clean samples for physical-property measurements, crystals from all three fractionation steps were etched with NaOH/HCl to remove the surface Al layer. To verify that the etching process did not alter the properties of \ce{CaMn_{2+$x$}Al_{10-$x$}}, magnetization measurements were performed on a step \#3 crystal as a representative example. \Cref{Fig:etching} shows the magnetic susceptibility of the step \#3 single crystal before and after etching. The overall temperature dependence of the susceptibility, including both the low-temperature upturn and the high-temperature behavior, remains essentially unchanged after etching.

\begin{figure}[t]
\centering
\includegraphics[width=\linewidth]{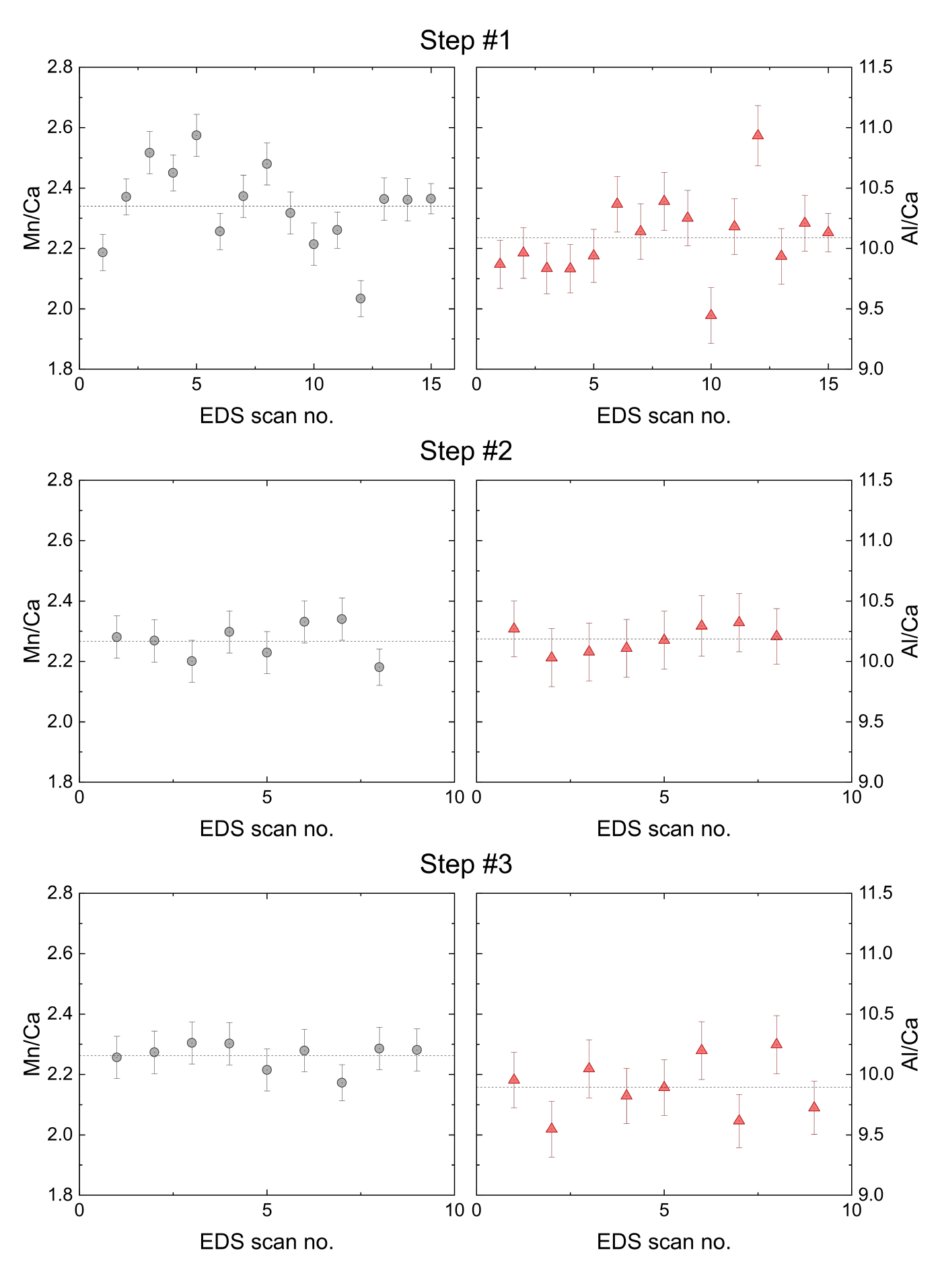}
\caption{
Mn/Ca (left panel) and Al/Ca (right panel) ratios, as determined from EDS area scans on the polished cross sections of rod-like \ce{CaMn_{2+$x$}Al_{10-$x$}} single crystals.
}
\label{Fig:EDS-ratios}
\end{figure}

\Cref{Fig:EDS-ratios} presents raw EDS data for \ce{CaMn_{2+$x$}Al_{10-$x$}}, plotted as Mn/Ca and Al/Ca ratios. The rod-like single crystals were cut perpendicular to the rod axis using a wire saw, and the cross sections were polished to collect EDS data. The Mn/Ca and Al/Ca ratios are relatively similar for crystals from steps \#2 and \#3, whereas the step \#1 crystals show a noticeably larger spread in the measured values.

\begin{figure}[htbp]
\centering
\includegraphics[width=\linewidth]{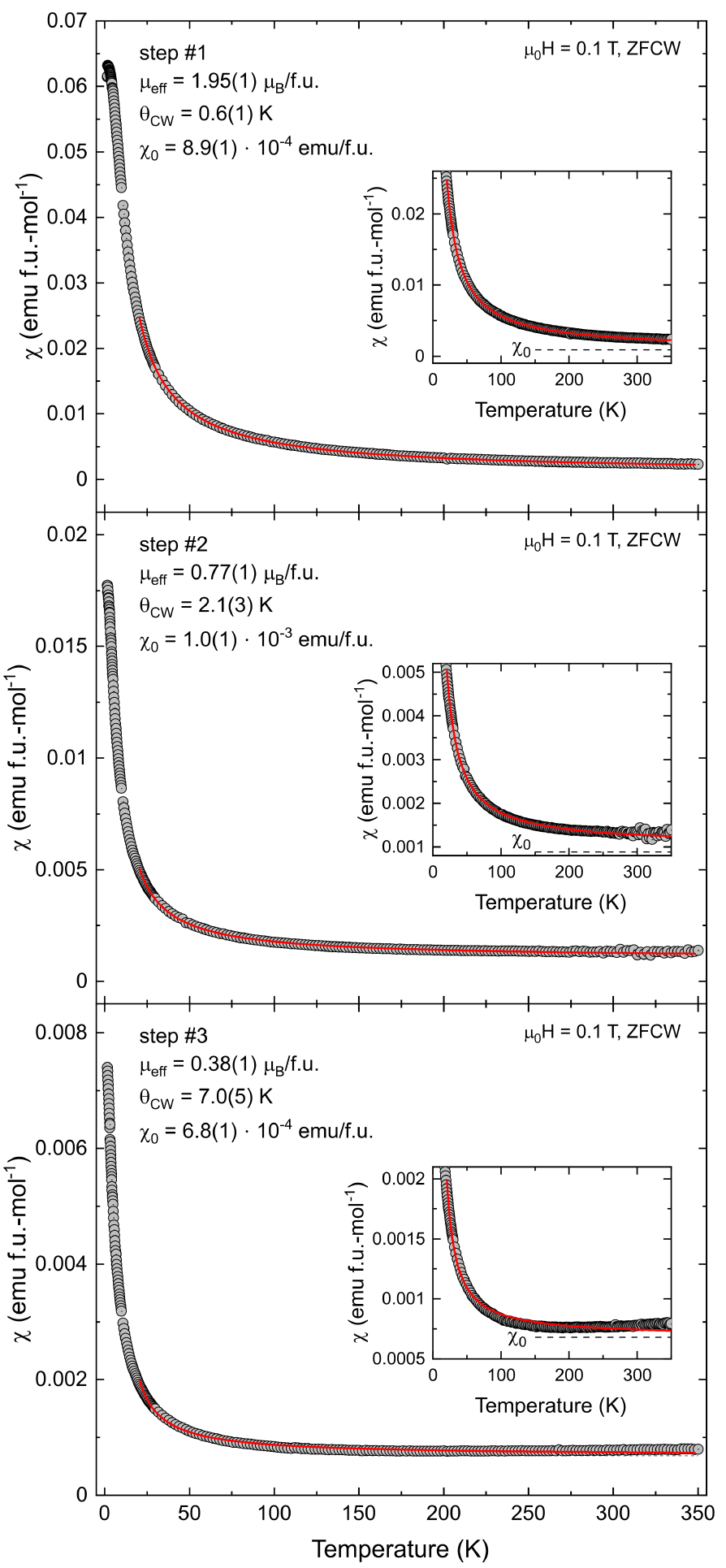}
\caption{
Temperature-dependent magnetic susceptibility of \ce{CaMn_{2+$x$}Al_{10-$x$}} single crystals, measured with magnetic field parallel to the $c$-axis. The Curie-Weiss fit (red) was performed in the 20 - 350 K temperature range. The inset shows a zoomed-in view of the susceptibility.
}
\label{Fig:CW-sanity}
\end{figure}

\Cref{Fig:CW-sanity} shows the magnetic susceptibility of \ce{CaMn_{2+$x$}Al_{10-$x$}} single crystals grown in steps \#1--3, together with Curie--Weiss fits over the 20--350 K temperature range. Below $T = 20$ K, the susceptibility begins to saturate; this range was excluded from the analysis. The suscpetibility of the step \#1 crystal is well described by the Curie--Weiss law over the entire 20-350 K temperature range. For step \#2, the fit quality is reasonable, with some deviation at high temperatures, while for step \#3 the fit quality is very poor, most probably due to temperature-dependent Pauli contribution [2]. To ensure consistency among all samples, the 20--100 K range was used for the analysis presented in \Cref{Fig:MPMS-CW}. The resulting effective magnetic moments determined from the fit over the 20--350 K range are slightly smaller than those obtained from the 20--100 K fits, except for step \#1, for which the value remains essentially unchanged.

\noindent \textbf{Reference} \\
\noindent[1] C. B. Shoemaker, D. A. Keszler, and D. P. Shoemaker,
Structure of $\mu$-MnAl4 with composition close to that of
quasicrystal phases, Acta Crystallographica Section B
Structural Science \textbf{45}, 13-20 (1989). \\
\noindent[2] P. Mohn, Magnetism in the Solid State: An Introduction, Springer Series in Solid-State Sciences, Vol. 134, Springer, Berlin/Heidelberg (2003), pp. 28–30

\end{document}